\documentclass[a4paper,fleqn,reqno]{amsart}
\usepackage{epic,eepic}
\usepackage{epsfig,cite,indentfirst}
\usepackage{multicol,boxedminipage}
\usepackage{varioref}
\begin{document}

\title[Elliptic height models and factorized DWPF]
{Two elliptic height models with factorized
 domain wall partition functions}

\author{O Foda, M Wheeler and M Zuparic}

\address{Department of Mathematics and Statistics,
         University of Melbourne,
         Parkville, Victoria 3010, Australia.}

\email{foda, mwheeler, mzup@ms.unimelb.edu.au}

\date{}

\newcommand{\field}[1]{\mathbb{#1}}
\newcommand{\CC}{\field{C}}
\newcommand{\NN}{\field{N}}
\newcommand{\ZZ}{\field{Z}}
\newcommand{\RR}{\field{R}}

\newcommand{\B}{{\mathcal B}}
\newcommand{\M}{{\mathcal M}}

\newcommand{\R}{{\mathcal R}}
\newcommand{\T}{{\mathcal T}}
\newcommand{\U}{{\mathcal U}}
\newcommand{\Y}{{\mathcal Y}}
\newcommand{\Z}{{\mathcal Z}}

\newcommand{\tT}{\widetilde{\mathcal T}}
\newcommand{\tU}{\widetilde{\mathcal U}}

\renewcommand{\P}{{\mathcal P}}

\def\nome{\textsf{q}}

\def\h{\textit{\scriptsize{h}}}
\def\o{\textit{\scriptsize{o}}}
\def\j{\textit{\scriptsize{j}}}
\def\k{\textit{\scriptsize{k}}}
\def\l{\textit{\scriptsize{l}}}
\def\m{\textit{\scriptsize{m}}}
\def\n{\textit{\scriptsize{n}}}
\def\p{\textit{\scriptsize{p}}}
\def\q{\textit{\scriptsize{q}}}
\def\r{\textit{\scriptsize{r}}}
\def\u{\textit{\scriptsize{u}}}
\def\v{\textit{\scriptsize{v}}}
\def\+{\textrm{\scriptsize{+}}}
\def\-{\textrm{\scriptsize{$-$}}}
\def\1{\textrm{\scriptsize{1}}}
\def\2{\textrm{\scriptsize{2}}}
\def\3{\textrm{\scriptsize{3}}}
\def\D{\textrm{\scriptsize{$\Delta$}}}

\def\ll{\left\lgroup}
\def\rr{\right\rgroup}

\newtheorem{ca}{Figure}
\newtheorem{corollary}{Corollary}
\newtheorem{definition}{Definition}
\newtheorem{example}{Example}
\newtheorem{lemma}{Lemma}
\newtheorem{notation}{Notation}
\newtheorem{proposition}{Proposition}
\newtheorem{remark}{Remark}
\newtheorem{theorem}{Theorem}
\newtheorem{property}{Property}
\def\ll{\left\lgroup}
\def\rr{\right\rgroup}
\newcommand{\Proof}{\medskip\noindent {\it Proof: }}
\def\proofend{\ensuremath{\square}}
\def\no{\nonumber}
\def\union{\mathop{\bigcup}}
\def\vac{|\mbox{vac}\rangle}
\def\cav{\langle\mbox{vac}|}
\def\lprod{\mathop{\prod{\mkern-29.5mu}{\mathbf\longleftarrow}}}
\def\rprod{\mathop{\prod{\mkern-28.0mu}{\mathbf\longrightarrow}}}
\def\a{\alpha}
\def\b{\beta}
\def\g{\gamma}
\def\d{\delta}
\def\e{\epsilon}
\def\s{\sigma}
\def\eps{\varepsilon}
\def\hb{\hat\beta}
\def\tg{\operatorname{tg}}
\def\ctg{\operatorname{ctg}}
\def\sh{\operatorname{sh}}
\def\ch{\operatorname{ch}}
\def\cth{\operatorname{cth}}
\def\th{\operatorname{th}}
\def\tla{\tilde{\lambda}}
\def\tmu{\tilde{\mu}}
\def\sul{\sum\limits}
\def\pl{\prod\limits}
\def\pd #1{\frac{\partial}{\partial #1}}
\def\ph #1{\phantom{                #1}}

\def\const{{\rm const}}
\def\argum{\{\mu_j\},\{\la_k\}}
\def\umarg{\{\la_k\},\{\mu_j\}}
\def\prodmu #1{\prod\limits_{j #1 k} \sinh(\mu_k-\mu_j)}
\def\prodla #1{\prod\limits_{j #1 k} \sinh(\lambda_k-\lambda_j)}
\newcommand{\bl}[1]{\makebox[#1em]{}}
\def\tr{\operatorname{tr}}
\def\Res{\operatorname{Res}}
\def\det{\operatorname{det}}
\newcommand{\boldN}{\boldsymbol{N}}
\newcommand{\bra}[1]{\langle\,#1\,|}
\newcommand{\ket}[1]{|\,#1\,\rangle}
\newcommand{\bracket}[1]{\langle\,#1\,\rangle}
\newcommand{\infinity}{\infty}
\renewcommand{\labelenumi}{\S\theenumi.}
\let\up=\uparrow
\let\down=\downarrow
\let\tend=\rightarrow
\hyphenation{boson-ic
             ferm-ion-ic
             para-ferm-ion-ic
             two-dim-ension-al
             two-dim-ension-al
             rep-resent-ative
             par-tition
             And-rews
             Gor-don
             boson-ic
             ferm-ion-ic
             para-ferm-ion-ic
             two-dim-ension-al
             two-dim-ension-al}
\renewcommand{\mod}{\textup{mod}\,}
\newcommand{\wt}{\text{wt}\,}

\begin{abstract}
We obtain factorized domain wall partition functions in
two elliptic height models:
{\bf 1.} A Felderhof-type model, which is new, and
{\bf 2.} A Perk-Schultz-type $gl(1|1)$ model of Deguchi
and Martin.
\end{abstract}

\maketitle

\setcounter{section}{-1}

\section{Introduction}

\subsection{Factorization in trigonometric vertex models} In
\cite{fwz}, we obtained factorized domain wall partition
functions (DWPF's)\footnote{For a review of previous results
on the subject, see \cite{korepin-review}.} in two series of
trigonometric vertex models:
{\bf 1.} The $N$-state Deguchi-Akutsu models, for
$N \in \{2, 3, 4\}$ (and conjectured the result for $N \geq 
5$), and
{\bf 2.} The $gl(r+1 | s+1)$ Perk-Schultz models, $\{r, s\} 
\in \NN$ (where given the symmetries of these models, the 
result is independent of $r$ and $s$).

\subsection{Asymmetry} These models were characterized by 
an asymmetry of the vertex weights under conjugation of 
state variables. For example, in the Deguchi-Akutsu model, 
with state variables $\sigma \in \{1, \cdots{}, N\}$, the 
vertex weights are non-invariant under the conjugation
$\sigma \rightarrow (N - \sigma + 1)$. In the Perk-Schultz
models, a similar property holds. Since one can trace the
factorization of the DWPF's obtained in \cite{fwz} to this
asymmetry, it is natural to look for height models\footnote{Also
known as face or interaction-round-face (IRF) models.}
with the same property.

\subsection{Factorization in elliptic height models} In this
work we consider height models, where the state variables are
heights, that live on the corners of the faces of a square
lattice. A weight is assigned to each face. The weights are
elliptic functions of the corresponding rapidities, external
fields (if any) and height variables. As in \cite{fwz}, the
models in this work are characterized by an asymmetry of the
weights, in the sense that the weights of certain vertices
(called line-permuting vertices) have different zeros, which
leads to the factorization of the DWPF's. 

\subsection{Summary of results} We obtain factorized DWPF's
for two elliptic height models:
{\bf 1.} A Felderhof-type model, which (to the best of our
knowledge) is new, and
{\bf 2.} A Perk-Schultz-type $gl(1|1)$ model of Deguchi
and Martin \cite{dm}. These are the first examples
of DWPF's for elliptic and/or height models.

\subsection{Outline of paper}
In Section {\bf 1}, we collect a number of basic definitions
related to elliptic height models to make the paper reasonably
self-contained.
In Section {\bf 2}, we introduce a new Felderhof-type elliptic
height model and obtain the corresponding factorized DWPF.
In Section {\bf 3}, we do the same for the $gl(1|1)$ elliptic
height model of Deguchi and Martin.
Section {\bf 4} contains brief remarks.
The presentation is elementary in the hope that the paper will 
be reasonably self-contained. 

\section{Height models}

\subsection{Faces and corners} We work on a square lattice, 
as in Figure {\bf \ref{lattice}}, with $L^{2}$ square faces 
$f_{ij}$, where $1 \leq i \leq L$ increases from top to 
bottom, and $1 \leq j \leq L$ increases from left to right. 
$f_{ij}$ has four corners that are labelled from top-left 
clockwise as 
$\{c_{i, j}, c_{i, j+1}, c_{i+1, j+1}, c_{i+1, j}\}$.

%
\begin{center}
\begin{minipage}{5.0in}
\setlength{\unitlength}{0.0016cm}
\begin{picture}(4800, 6000)(-2500, -1500)
\thicklines
%
%
\path(0000,3600)(3600,3600)
\path(0000,3000)(3600,3000)
\path(0000,2400)(3600,2400)
\path(0000,1800)(3600,1800)
\path(0000,1200)(3600,1200)
\path(0000,0600)(3600,0600)
\path(0000,0000)(3600,0000)
%
%
\path(0000,3600)(0000,0000)
\path(0600,3600)(0600,0000)
\path(1200,3600)(1200,0000)
\path(1800,3600)(1800,0000)
\path(2400,3600)(2400,0000)
\path(3000,3600)(3000,0000)
\path(3600,3600)(3600,0000)
\put(0150,3250){$f_{11}$}
\put(0150,2650){$f_{21}$}
\put(3150,3250){$f_{1L}$}
\put(0150,0250){$f_{L1}$}

\put(0750,0250){$f_{L2}$}
\put(3150,0250){$f_{LL}$}
\put(-400,3600){$h_{00}$}
\put(-400,3000){$h_{10}$}
\put(-400,0000){$h_{L0}$}

\put(0000,3600){\blacken\circle{50,50}}
\put(3600,3600){\blacken\circle{50,50}}
\put(0000,3000){\blacken\circle{50,50}}
\put(0000,2400){\blacken\circle{50,50}}
\put(0000,1800){\blacken\circle{50,50}}
\put(0000,1200){\blacken\circle{50,50}}
\put(0000,0600){\blacken\circle{50,50}}
\put(3700,3600){$h_{0L}$}
\put(3700,0000){$h_{LL}$}
\put(0000,0000){\blacken\circle{50,50}}
\put(0600,0000){\blacken\circle{50,50}}
\put(0500,-300){$h_{10}$}
\put(3600,0000){\blacken\circle{50,50}}
\thinlines
%
%
\dashline{60.00}(-200,3300)(0050,3300)
\dashline{60.00}(0550,3300)(3050,3300)
\dashline{60.00}(3550,3300)(3900,3300)

\dashline{60.00}(-200,2700)(0050,2700)
\dashline{60.00}(0550,2700)(3900,2700)
\thicklines
\whiten\path(-500,3255)(-320,3300)(-500,3345)(-500,3255)
\path(-500,3300)(-680,3300)
\thinlines
\put(-1400,3300){$u_1, p_1$}
\thicklines
\whiten\path(-500,2655)(-320,2700)(-500,2745)(-500,2655)
\path(-500,2700)(-680,2700)
\thinlines
\dashline{60.00}(-200,2100)(3900,2100)
\thicklines
\whiten\path(-500,2055)(-320,2100)(-500,2145)(-500,2055)
\path(-500,2100)(-680,2100)
\thinlines
\dashline{60.00}(-200,1500)(3900,1500)
\thicklines
\whiten\path(-500,1455)(-320,1500)(-500,1545)(-500,1455)
\path(-500,1500)(-680,1500)
\thinlines
\dashline{60.00}(-200,0900)(3900,0900)
\thicklines
\whiten\path(-500,0855)(-320,0900)(-500,0945)(-500,0855)
\path(-500,0900)(-680,0900)
\thinlines
\put(-1400,2700){$u_2, p_2$}
\dashline{60.00}(-200,0300)(0050,0300)
\dashline{60.00}(0550,0300)(0650,0300)
\dashline{60.00}(1150,0300)(3050,0300)
\dashline{60.00}(3550,0300)(3900,0300)
\thicklines
%
\whiten\path(-500,0255)(-320,0300)(-500,0345)(-500,0255)
\path(-500,0300)(-680,0300)
\thinlines
\put(-1400,0300){$u_L, p_L$}
%
\dashline{60.00}(0300,3900)(0300,3550)
\dashline{60.00}(0300,3050)(0300,2900)
\dashline{60.00}(0300,2500)(0300,0550)
\dashline{60.00}(0300,0050)(0300,-200)
\thicklines
\whiten\path(0255,-500)(0345,-500)(0300,-320)(0255,-500)
\path(0300,-680)(0300,-500)
\thinlines
\put(0200,-0900){$v_1,$}
\put(0200,-1200){$q_1$}
\dashline{60.00}(0900,3900)(0900,0550)
\dashline{60.00}(0900,0050)(0900,-200)
\thicklines
\whiten\path(0855,-500)(0945,-500)(0900,-320)(0855,-500)
\path(0900,-680)(0900,-500)
\thinlines
\put(0800,-0900){$v_2,$}
\put(0800,-1200){$q_2$}
\dashline{60.00}(1500,3900)(1500,-200)
\thicklines
\whiten\path(1455,-500)(1545,-500)(1500,-320)(1455,-500)
\path(1500,-680)(1500,-500)
\thinlines
\dashline{60.00}(2100,3900)(2100,-200)
\thicklines
\whiten\path(2055,-500)(2145,-500)(2100,-320)(2055,-500)
\path(2100,-680)(2100,-500)
\thinlines
\dashline{60.00}(2700,3900)(2700,-200)
\thicklines
\whiten\path(2655,-500)(2745,-500)(2700,-320)(2655,-500)
\path(2700,-680)(2700,-500)
\thinlines
\dashline{60.00}(3300,3900)(3300,3550)
\dashline{60.00}(3300,3050)(3300,0550)
\dashline{60.00}(3300,0050)(3300,-200)
\thicklines
\whiten\path(3255,-500)(3345,-500)(3300,-320)(3255,-500)
\path(3300,-680)(3300,-500)
\thinlines
\put(3200,-0900){$v_L,$}
\put(3200,-1200){$q_L$}
\end{picture}
\begin{ca}
\label{lattice}
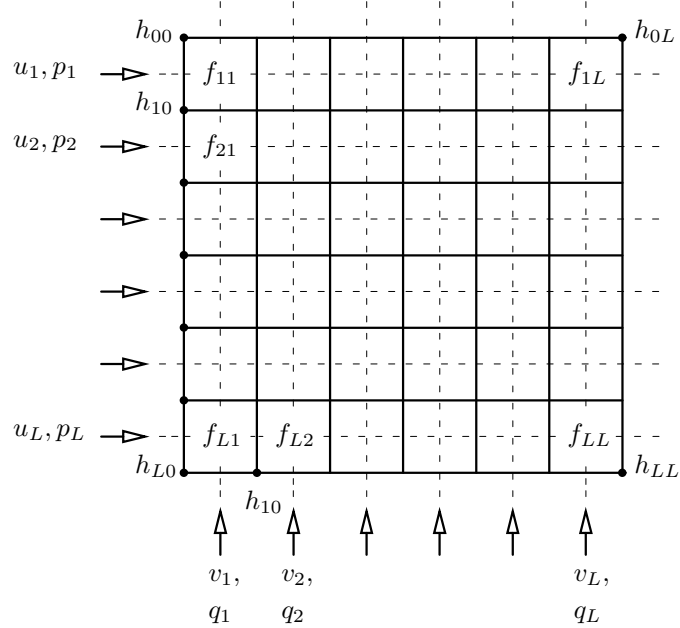
A square lattice with $L^{2}$ faces $f_{ij}$. Rapidities
$\{u, v\}$ and external fields $\{p, q\}$ flow along lines
that cross the faces. Height variables $h_{ij}$ live on the
corners.
\end{ca}
\end{minipage}
\end{center}

\bigskip

\subsection{Heights and restrictions}

We assign each corner $c_{ij}$ a height variable $h_{ij}$,
$0 \leq i, j \leq L$. In height models such as Baxter's
solid-on-solid model \cite{baxter-book}, the heights are
integral (possibly up to an overall shift). In the model
of Section {\bf 2}, the height variables depend linearly
on the external fields which are continuous parameters,
so they are no longer integral. We define the heights and
the restrictions that they obey on a model by model basis
in Sections {\bf 2} and {\bf 3}.

\subsection{Flow lines, orientations and variables} There are
$L$ horizontal and $L$ vertical lines that intersect at the middle
points of $f_{ij}$. They indicate the flow of rapidities and
external fields through $f_{ij}$.
We assign the $i$-th horizontal line an orientation from left
to right, a complex rapidity $u_i$ and a complex external
field $p_i$. We assign the $j$-th vertical line
an orientation from bottom to top, a complex rapidity 
$v_j$ and a complex external field $q_j$, as in
Figure {\bf \ref{lattice}}.

\subsection{Weights and Yang-Baxter equations}

We assign each $f_{ij}$ a weight $w_{ij}$
that depends on the height variables on its corners, the difference
of the rapidity variables and the two external field variables (if
any) flowing through it. The weights satisfy a set of Yang-Baxter
equations. The weights and Yang-Baxter equations of the models
discussed in this paper are given in Sections {\bf 2} and {\bf 3}.

\subsection{Elliptic functions and a theorem} 
Following the conventions used in \cite{baxter-book}, Chapter 
{\bf 15}, we consider the elliptic function

\begin{equation}
H(u) =
2 \nome^{1/4} 
\sin \ll \frac{\pi u}{2I}\rr
\prod_{n=1}^{\infty}
\ll
1 - 2 \nome^{2n} \cos \frac{\pi u}{I} + \nome^{4n}
\rr 
\ll 
1 - \nome^{2n}
\rr
\label{H}
\end{equation}

\noindent where $u \in \CC$, 
\nome\ $= \exp\ll \frac{- \pi I'}{I} \rr$, $2I$ and $2I'$ (usually
called $2K$ and $2K'$) are respectively the (real) width and height 
of an (upright) rectangle $R$ in the complex $u$-plane, so that
$0 < \nome < 1$. It is convenient to define

\begin{eqnarray}
[u] = \frac{H(u)}{2 \nome^{1/4}}
\label{HH}
\end{eqnarray}

\noindent which is entire and satisfies the quasi-periodicity
properties

\begin{eqnarray}
&&
[u + 2I] = -[u] \\ 
\nonumber \\
&&
[u + 2iI']
=
- \frac{1}{\nome} \exp\ll \frac{- \pi i u}{I}\rr [u]
\end{eqnarray}

\subsection*{Theorem 1} If $f(u)$ is an entire function that
satisfies the quasi-periodicity conditions

\begin{equation}
f(u + 2I)
=
(-)^L f(u)
\end{equation}

\begin{equation}
f(u + 2iI')
=
\ll 
\frac{-1}{\nome}
\rr^{L}
\exp
\ll 
\frac{-\pi i (L u - \eta)}{I}
\rr
f(u)
\end{equation}

\noindent then

\begin{eqnarray}
f(u)
=
\kappa
\ll
\prod_{j=1}^{L-1}
[u-\zeta_j]
\rr
[u-\eta+\sum_{j=1}^{L-1}\zeta_j]
\end{eqnarray}
where $\kappa$ and $\zeta_1,\ldots,\zeta_{L-1}$ are constants.

\bigskip

\noindent {\it Proof.} This is a refinement of Theorem {\bf 15(c)} 
of \cite{baxter-book}, and the proof uses a similar argument. 
Choose the period rectangle $R$ such that $f(u)$ has no zeros on 
the boundary $\partial R$, and integrate $\frac{f'(u)}{f(u)}$ 
on the anti-clockwise contour $\partial R$. From the quasi-periodicity 
conditions it follows that

\begin{eqnarray}
\oint_{\partial R} \frac{f'(u)}{f(u)}du=2\pi iL
\end{eqnarray}

Hence the sum of residues of $\frac{f'(u)}{f(u)}$ in $R$ is equal
to $L$, showing $f(u)$ has exactly $L$ zeros in $R$ (counting a
zero of order $n$ with multiplicity $n$). Writing the $L$ zeros
as $\zeta_1,\ldots,\zeta_L$, we define the function
$\phi(u)=\prod_{j=1}^{L}[u-\zeta_j]$. Since
$\frac{d}{du} \log (f(u)/\phi(u))$ is doubly periodic
(by construction) and holomorphic (also by construction) one has

\begin{eqnarray}
\frac{d}{du} \log
\ll \frac{f(u)}{\phi(u)} \rr = \lambda
\end{eqnarray}

\noindent where $\lambda$ is a constant. Integrating, we obtain
$f(u)=\kappa e^{\lambda u}\prod_{j=1}^{L}[u-\zeta_j]$. Using the
quasi-periodicity conditions of $f(u)$, we can, without loss of
generality, choose 
$\lambda=0$ and $\zeta_L=\eta-\sum_{j=1}^{L-1}\zeta_j$, which
concludes the proof.

\section{A Felderhof-type height model}

In this section, we introduce an elliptic height model with weights that
depend on rapidities, external fields and height variables. In the
trigonometric limit, it reduces to the first in a series of models
introduced by Deguchi and Akutsu in \cite{da2}. In that same limit,
and decoupling the dependence on the heights\footnote{This can be
achieved, for example, by introducing a parameter $\xi \in i\RR$,
shifting all height variables by $\xi$ (the Yang-Baxter equations
remain satisfied), then taking the limit $\xi \rightarrow i\infty$.},
it reduces to the trigonometric limit of the elliptic Felderhof
vertex model, which is the 2-state Deguchi-Akutsu model \cite{da1}.

\subsection{Notation}

Given the rapidities $\{ u, v\} \in \CC$, external fields
$\{ p, q\} \in \CC$, an upper-left corner height $h \in \CC$,
$\{\Delta_1, \Delta_2\} \in \{0, 1   \}$, 
$  \Delta_3             \in \{0, 1, 2\}$, 
we use the notation

\begin{eqnarray}
W_{uv}
\ll
\begin{array}{ll}
\h                          & \h \+ \q      \- \D_{\1}  \\
\h     \+ \p \- \D_{\2}    & \h \+ \q \+ \p \- \D_{\3} 
\end{array}
\rr
\end{eqnarray}

\noindent for the weight assigned to the 
vertex\footnote{In the sequel, we simply 
say `vertex' instead of `face configuration'.}
represented in Figure {\bf \ref{general-vertex-f}}.

%
\begin{center}
\begin{minipage}{3.0in}
\setlength{\unitlength}{0.0016cm}
\begin{picture}(3000,3000)(0000,200)
\thicklines
\path(1500,2400)(2700,2400)
\path(1500,1200)(2700,1200)
%
\path(1500,2400)(1500,1200)
\path(2700,2400)(2700,1200)
%
%
\thinlines
\dashline{60.00}(1200,1800)(3000,1800)
\thicklines
%
\whiten\path(0840,1740)(1000,1800)(0840,1860)(0840,1740)
\path(0680,1800)(0840,1800)
 \put(0500,1900){\u\scriptsize{,}\p}
%
%
\thinlines
\dashline{60.00}(2100,2700)(2100,0900)
\thicklines
%
\whiten\path(2040,0540)(2100,0700)(2160,0540)(2040,0540)
%
\path(2100,0540)(2100,0380)
\put(2200,0380){\v\scriptsize{,}\q}
\put(1500,2400){\blacken\ellipse{050}{050}}
\put(1300,2500){\h}
\put(1500,1200){\blacken\ellipse{050}{050}}
\put(1300,1000){\h\+\p\scriptsize{$-$}$\D_{\2}$}
\put(2700,2400){\blacken\ellipse{050}{050}}
\put(2500,2500){\h\+\q\scriptsize{$-$}$\D_{\1}$}
\put(2700,1200){\blacken\ellipse{050}{050}}
\put(2500,1000){\h\+\q\+\p\scriptsize{$-$}$\D_{\3}$}
\end{picture}
\begin{ca}
\label{general-vertex-f}
A Felderhof-type face configuration.
\end{ca}
\end{minipage}
\end{center}
\bigskip

\subsection{Height restrictions} 
For $p = q = \frac{1}{2}$, we require that the heights 
satisfy the same restriction as in Baxter's solid-on-solid
model \cite{baxter-book}, up to a normalization. More 
precisely,

\begin{equation}
h_{i, j} - h_{i+1, j} = \pm \frac{1}{2}, \quad 
h_{i, j} - h_{i, j+1} = \pm \frac{1}{2}
\end{equation}

\subsection{The crossing parameter $= \frac{1}{2}$} The vertex 
weights will be parametrized in terms of the elliptic functions 
$[u]$. As defined in Equations {\bf \ref{H}} and {\bf \ref{HH}}, 
$[u]$ depends on the real parameters, $I$ and $I'$, which are 
the magnitudes of the half-periods of $[u]$. In the Felderhof-type 
model discussed in this section, we set $I = 1$\footnote{In the limit 
of zero external fields, that is $p=q=\frac{1}{2}$, this is equivalent 
to setting {\it the crossing parameter} in Baxter's solid-on-solid model 
to the free fermion point. For details, see \cite{baxter-book}}.

\subsection{The weights}

In the above notation, the non-zero weights are

\begin{eqnarray}
W_{uv}
\ll
\begin{array}{ll}
\h                           & \h\+ \q                 \\
\h     \+ \p                 & \h\+ \q     \+ \p
\end{array}
\rr
&=& a_{+}(u,v,p,q) = [u-v+p+q]
\\
W_{uv}
\ll
\begin{array}{ll}
\h                        & \h\+ \q  \- \1            \\
\h     \+ \p      \- \1   & \h\+ \q  \+ \p     \- \2
\end{array}
\rr
&=& a_{-}(u,v,p,q) = [v-u+p+q]
\end{eqnarray}
\begin{eqnarray}
W_{uv}
\ll
\begin{array}{ll}
\h                        & \h \+ \q \- \1            \\
\h     \+ \p              & \h \+ \q \+ \p     \- \1
\end{array}
\rr
&=&
b_{+}(u,v,p,q,h)
\\
&=&
\frac{
[2h]^\frac{1}{2}
[2(h+p+q)]^\frac{1}{2}
}
{
[2(h+p)]^\frac{1}{2}
[2(h+q)]^\frac{1}{2}
}
[u-v+q-p]
\nonumber
\end{eqnarray}
\begin{eqnarray}
W_{uv}
\ll
\begin{array}{ll}
\h                        & \h \+\q                \\
\h     \+ \p \- \1        & \h \+\q     \+\p \- \1
\end{array}
\rr
&=&
b_{-}(u,v,p,q,h)
\\
&=&
\frac{
[2h]^\frac{1}{2}
[2(h+p+q)]^\frac{1}{2}
}
{
[2(h+p)]^\frac{1}{2}
[2(h+q)]^\frac{1}{2}
}
[u-v+p-q]
\nonumber
\end{eqnarray}
\begin{eqnarray}
W_{uv}
\ll
\begin{array}{ll}
\h           & \h \+\q                \\
\h     \+\p  & \h \+\q     \+\p \- \1
\end{array}
\rr
&=&
c_{+}(u,v,p,q,h)
\\
&=&
\frac{
[2p]^\frac{1}{2}
[2q]^\frac{1}{2}
}
{
[2(h+p)]^\frac{1}{2}
[2(h+q)]^\frac{1}{2}
}
[v-u+p+q+2h]
\nonumber
\end{eqnarray}
\begin{eqnarray}
W_{uv}
\ll
\begin{array}{ll}
\h              & \h\+\q  \- \1           \\
\h\+\p \- \1      & \h\+\q\+\p \- \1
\end{array}
\rr
&=&
c_{-}(u,v,p,q,h)
\\
&=&
\frac{
[2p]^\frac{1}{2}
[2q]^\frac{1}{2}
}
{
[2(h+p)]^\frac{1}{2}
[2(h+q)]^\frac{1}{2}
}
[u-v+p+q+2h]
\nonumber
\end{eqnarray}

\subsection{The Yang-Baxter equations.}

For rapidities $\{u,v,w\}$, external fields $\{p,q,r\}$, 
and non-negative integers $\{k,l,m,n,o\}$, the above 
weights satisfy the Yang-Baxter equations

\begin{eqnarray}
&\displaystyle{\sum_{j \geq 0}}&
W_{uv}
\ll
\begin{array}{ll}
\h        & \h\+\q\-\j \\
\h\+\p\-\o& \h\+\q\+\p\-\n
\end{array}
\rr
W_{uw}
\ll
\begin{array}{ll}
\h\+\q\-\j     & \h\+\q\+\r\-\l     \\
\h\+\q\+\p\-\n & \h\+\q\+\r\+\p\-\m
\end{array}
\rr
\nonumber \\
&\times&
W_{vw}
\ll
\begin{array}{ll}
\h        & \h\+\r\-\k        \\
\h\+\q\-\j& \h\+\r\+\q\-\l
\end{array}
\rr
\nonumber \\
&=&
\nonumber \\
&\displaystyle{\sum_{j \geq 0}}&
W_{uv}
\ll
\begin{array}{ll}
\h\+\r\-\k    & \h\+\r\+\q\-\l     \\
\h\+\r\+\p\-\j& \h\+\r\+\q\+\p\-\m
\end{array}
\rr
W_{uw}
\ll
\begin{array}{ll}
\h        & \h\+\r \- \k     \\
\h\+\p \- \o& \h\+\r\+\p \- \j
\end{array}
\rr
\nonumber \\
&\times&
W_{vw}
\ll
\begin{array}{ll}
\h\+\p \- \o    & \h\+\p\+\r \- \j     \\
\h\+\p\+\q \- \n& \h\+\p\+\r\+\q \- \m
\end{array}
\rr
\end{eqnarray}

\noindent {\it Proof.} This can be proved by direct computation
using elliptic function identities, along the same lines as in
\cite{baxter-book}. For example, when
$\{k,l,m,n,o\} = \{0,1,1,1,1\}$, the Yang-Baxter equation is

\begin{eqnarray}
&\displaystyle{\sum_{j = 0}^{1}}&
W_{uv}
\ll
\begin{array}{ll}
\h        & \h\+\q\-\j     \\
\h\+\p\-\1& \h\+\q\+\p\-\1
\end{array}
\rr
W_{uw}
\ll
\begin{array}{ll}
\h\+\q\-\j    & \h\+\q\+\r\-\1     \\
\h\+\q\+\p\-\1& \h\+\q\+\r\+\p\-\1
\end{array}
\rr
\nonumber \\
&\times&
W_{vw}
\ll
\begin{array}{ll}
\h        & \h\+\r          \\
\h\+\q\-\j& \h\+\r\+\q\-\1
\end{array}
\rr
\nonumber \\
&=& W_{uv}
\ll
\begin{array}{ll}
\h\+\r        & \h\+\r\+\q\-\1      \\
\h\+\r\+\p\-\1& \h\+\r\+\q\+\p\-\1
\end{array}
\rr
W_{uw}
\ll
\begin{array}{ll}
\h        & \h\+\r          \\
\h\+\p\-\1& \h\+\r\+\p\-\1
\end{array}
\rr
\nonumber \\
&\times&
W_{vw}
\ll
\begin{array}{ll}
\h\+\p\-\1    & \h\+\p\+\r\-\1      \\
\h\+\p\+\q\-\1& \h\+\p\+\r\+\q\-\1
\end{array}
\rr
\end{eqnarray}

Using the expressions for the weights, we obtain

\begin{eqnarray}
&&
c_{-}(u,v,p,q,h)\ a_{+}(u,w,p,r)\ b_{-}(v,w,q,r,h)
\nonumber \\
&+&
b_{-}(u,v,p,q,h)\ c_{-}(u,w,p,r,h+q)\ c_{+}(v,w,q,r,h)
\nonumber \\
&=&
c_{-}(u,v,p,q,h+r)\ b_{-}(u,w,p,r,h)\ a_{+}(v,w,q,r)
\end{eqnarray}

Writing the weights in terms of elliptic functions, one can 
eliminate common factors, and the proof of the equation
reduces to proving

\begin{eqnarray}
&\ph{\times}&
[u-v+p+q+2h] [u-w+p+r] [v-w+q-r] [2(h+q+r)]
\nonumber \\
&+&
[u-v+p-q] [u-w+p+r+2(h+q)] [w-v+q+r+2h] [2r]
\nonumber \\
&=&
[u-v+p+q+2(h+r)] [u-w+p-r] [v-w+q+r] [2(h+q)]
\end{eqnarray}

\noindent which proceeds by noting that the ratio of the
left-hand-side and right-hand-side is doubly periodic and
entire in $u$, and therefore a constant with respect to $u$.
Setting $u=v-p+q$, the constant is found to be 1.

\subsection{Switching off the external fields} Setting $p 
= q = \frac{1}{2}$ is equivalent to switching off the external 
fields. This becomes clear by inspection of the vertex weights, 
which up to normalization become equal to those of Baxter's 
solid-on-solid model at the free fermion point.

\subsection{The external fields tilt the heights} One can 
think of the external fields $p \neq \frac{1}{2}$ and/or 
$q \neq \frac{1}{2}$, as effectively {\it tilting} the heights 
of the lattice faces that they flow through. This tilt is with 
respect to the line along which a field flows. This effectively 
adds to or subtracts from the height differences that are the case 
in the absence of external fields.

\subsection{The $c_{+}$ vertex}

In discussions of DWBC's and DWPF's, the $c_{+}$ vertex,
see Figure {\bf \ref{c+}}, plays a special role: It is
the DWPF for a $1\times 1$ square lattice.

%
\begin{center}
\begin{minipage}{3.0in}
\setlength{\unitlength}{0.0016cm}
\begin{picture}(3000,3000)(0,0200)
\thicklines
%
\path(1500,2400)(2700,2400)
\path(1500,1200)(2700,1200)
%
\path(1500,2400)(1500,1200)
\path(2700,2400)(2700,1200)
\thinlines
\dashline{60.00}(1200,1800)(3000,1800)
\thicklines
%
\whiten\path(0840,1740)(1000,1800)(0840,1860)(0840,1740)
\path(0680,1800)(0840,1800)
 \put(0500,1900){\u, \p}
\thinlines
\dashline{60.00}(2100,2700)(2100,0900)
\thicklines
%
\whiten\path(2040,0540)(2100,0700)(2160,0540)(2040,0540)
%
\path(2100,0540)(2100,0380)
\put(2200,0380){\v,\q}
\put(1500,2400){\blacken\ellipse{050}{050}}
\put(1300,2500){\h}
\put(1500,1200){\blacken\ellipse{050}{050}}
\put(1300,1000){\h\+\p}
\put(2700,2400){\blacken\ellipse{050}{050}}
\put(2500,2500){\h\+\q}
\put(2700,1200){\blacken\ellipse{050}{050}}
\put(2500,1000){\h\+\q\+\p\textit{\scriptsize{--}}\1}
\end{picture}
\begin{ca}
\label{c+}
The Felderhof-type $c_{+}$ vertex.
\end{ca}
\end{minipage}
\end{center}
\bigskip

\subsection{Domain wall boundary conditions (DWBC)}

We define the DWBC's as an {\it expanded} $c_{+}$ vertex,
as in Figure {\bf \ref{dwbcf}}: Given the external fields 
$\{p, q\}$ and starting from $h_{00} = h$ at the top-left 
corner, the boundary heights change by
$q_j $ from left to right along the upper boundary,
$p_i-1$ from top to bottom along the right boundary,
$-q_j+1$ from right to left along the lower boundary, 
and $-p_i$ from bottom to top along the left boundary.

%
\begin{center}
\begin{minipage}{5.0in}
\setlength{\unitlength}{0.0016cm}
\begin{picture}(4800, 6000)(-2400, -1500)
%
\thicklines
%
%
\path(0000,3600)(3600,3600)
\path(0000,3000)(3600,3000)
\path(0000,2400)(3600,2400)
\path(0000,1800)(3600,1800)
\path(0000,1200)(3600,1200)
\path(0000,0600)(3600,0600)
\path(0000,0000)(3600,0000)
%
%
\path(0000,3600)(0000,0000)
\path(0600,3600)(0600,0000)
\path(1200,3600)(1200,0000)
\path(1800,3600)(1800,0000)
\path(2400,3600)(2400,0000)
\path(3000,3600)(3000,0000)
\path(3600,3600)(3600,0000)
\put(-0200,3700){$h          $}
\put(-0900,3000){$h + p_{1  }$}
\put(-0900,2400){$h + p_{1,2}$}
\put(-0900, 000){$h + p_{1,L}$}

\put(3700,3700){$h + q_{1, L}              $}
\put(3700,3000){$h + q_{1, L} + p_{1  } - 1$}
\put(3700,2400){$h + q_{1, L} + p_{1,2} - 2$}
\put(3700,0000){$h + q_{1, L} + p_{1,L} - L$}

\put(0000,3600){\blacken\circle{50,50}}
\put(0000,3000){\blacken\circle{50,50}}
\put(0000,2400){\blacken\circle{50,50}}
\put(0000,1800){\blacken\circle{50,50}}
\put(0000,1200){\blacken\circle{50,50}}
\put(0000,0600){\blacken\circle{50,50}}

\put(3600,3600){\blacken\circle{50,50}}

\put(0000,0000){\blacken\circle{50,50}}
\put(0600,0000){\blacken\circle{50,50}}
\put(3600,0000){\blacken\circle{50,50}}
\thinlines
%
%
\dashline{60.00}(-200,3300)(3900,3300)
\dashline{60.00}(-200,2700)(3900,2700)
\dashline{60.00}(-200,2100)(3900,2100)
\dashline{60.00}(-200,1500)(3900,1500)
\dashline{60.00}(-200,0900)(3900,0900)
\dashline{60.00}(-200,0300)(3900,0300)
%
%
\dashline{60.00}(0300,3900)(0300,-200)

\thicklines
\blacken\path(0555,-500)(0645,-500)(0600,-320)(0555,-500)
\path(0600,-980)(0600,-500)
\put(0600,-1200){$h + q_1 + p_{1, L} -1$}

\thicklines
\blacken\path(1155,-500)(1245,-500)(1200,-320)(1155,-500)
\path(1200,-680)(1200,-500)
\put(1200,-0900){$h + q_{1,2} + p_{1, L} -2$}

\thinlines
\dashline{60.00}(0900,3900)(0900,-200)
\dashline{60.00}(1500,3900)(1500,-200)
\dashline{60.00}(2100,3900)(2100,-200)
\dashline{60.00}(2700,3900)(2700,-200)
\dashline{60.00}(3300,3900)(3300,-200)
\end{picture}
\begin{ca}
\label{dwbcf}
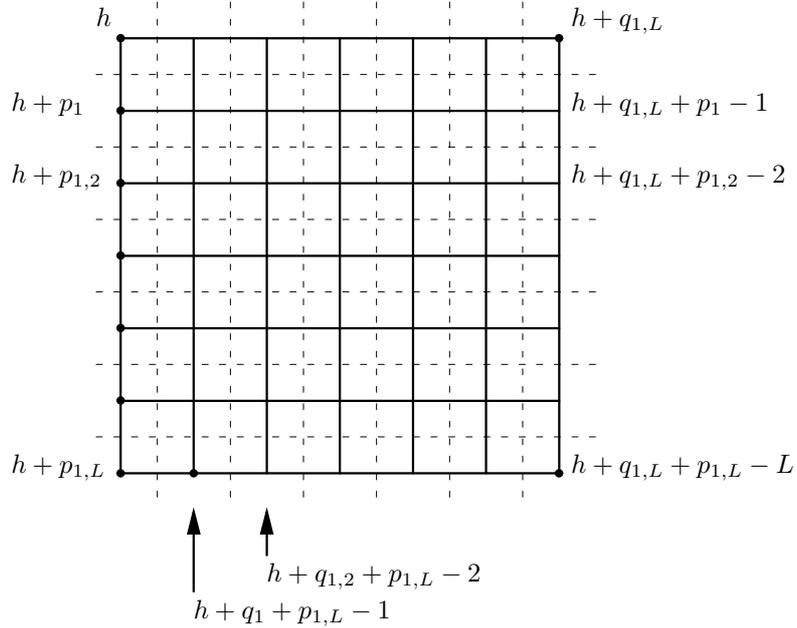
Felderhof-type height domain wall boundary conditions.
We use the notation $p_{i, j} = \sum_{k=i}^{j} p_k$,
{\it etc.}
\end{ca}
\end{minipage}
\end{center}
\bigskip

\subsection{Domain wall partition function (DWPF)}

The DWPF on an $L\times L$ lattice, $Z_{L\times L}$, is the
sum over all weighted configurations that satisfy the DWBC.
The weight of each configuration is the product of the weights
of the vertices

\begin{eqnarray}
Z^{}_{L\times L}
=
\sul_{{\rm configurations}}
\ll
\pl_{\rm vertices} w_{ij}
\rr
\label{physical}
\end{eqnarray}

\subsection{Line permuting vertices}
In proofs of DWPF's two vertices play an important role.
These are the $a$-type vertices which can be used to
permute adjacent flow lines.

%
\begin{center}
\begin{minipage}{5.0in}
\setlength{\unitlength}{0.0016cm}
\begin{picture}(3000,3000)(-200,0)
\thicklines
%
\path(1500,2400)(2700,2400)
\path(1500,1200)(2700,1200)
%
\path(1500,2400)(1500,1200)
\path(2700,2400)(2700,1200)
\thinlines
\dashline{60.00}(1200,1800)(3000,1800)
\thicklines
%
\whiten\path(0840,1740)(1000,1800)(0840,1860)(0840,1740)
\path(0680,1800)(0840,1800)
 \put(0500,1900){\u, \p}
\thinlines
\dashline{60.00}(2100,2700)(2100,0900)
\thicklines
%
\whiten\path(2040,0540)(2100,0700)(2160,0540)(2040,0540)
%
\path(2100,0540)(2100,0380)
\put(2200,0380){\v,\q}
\put(1500,2400){\blacken\ellipse{050}{050}}
\put(1300,2500){\h}
\put(1500,1200){\blacken\ellipse{050}{050}}
\put(1300,1000){\h\+\p}
\put(2700,2400){\blacken\ellipse{050}{050}}
\put(2500,2500){\h\+\q}
\put(2700,1200){\blacken\ellipse{050}{050}}
\put(2500,1000){\h\+\q\+\p}
%
%
\path(4500,2400)(5700,2400)
\path(4500,1200)(5700,1200)
%
\path(4500,2400)(4500,1200)
\path(5700,2400)(5700,1200)
%
\thinlines
\dashline{60.00}(4200,1800)(6000,1800)
\thicklines
%
\whiten\path(3840,1740)(4000,1800)(3840,1860)(3840,1740)
\path(3680,1800)(3840,1800)
 \put(3500,1900){\u, \p}
%
\thinlines
\dashline{60.00}(5100,2700)(5100,0900)
\thicklines
%
\whiten\path(5040,0540)(5100,0700)(5160,0540)(5040,0540)
%
\path(5100,0540)(5100,0380)
 \put(5200,0380){\v,\q}
\put(4500,2400){\blacken\ellipse{050}{050}}
\put(4300,2500){\h}
\put(4500,1200){\blacken\ellipse{050}{050}}
\put(4300,1000){\h\+\p\textit{\scriptsize{--}}\1}
\put(5700,2400){\blacken\ellipse{050}{050}}
\put(5500,2500){\h\+\q\textit{\scriptsize{--}}\1}
\put(5700,1200){\blacken\ellipse{050}{050}}
\put(5500,1000){\h\+\q\+\p\textit{\scriptsize{--}}\2}
\end{picture}
\begin{ca}
\label{face}
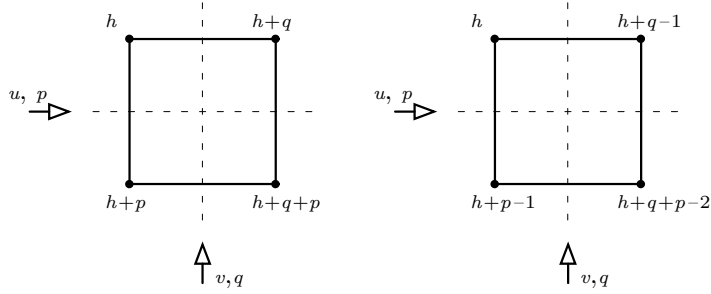
The Felderhof-type line permuting vertices $a_{+}$ 
and $a_{-}$.
\end{ca}
\end{minipage}
\end{center}
\bigskip

\subsection{Different zeros} The weights of the line permuting 
vertices, $[u-v+p+q]$ and $[v-u+p+q]$, have different zeros. 
This is the property that will allow us to obtain the zeros of 
the DWPF and compute it in factorized form.

\subsection{Properties of the partition function} The following
four properties determine the partition function uniquely.

\subsubsection{Property 1: Quasi-periodicity} The partition
function is entire in $u_1$ and satisfies the quasi-periodicity
conditions

\begin{equation}
Z_{L\times L}
\ll
u_1+2, \ldots, u_L, \{v\}, \{p\}, \{q\}, h
\rr 
= 
(-)^L Z_{L\times L}
\ll
\{u\},\{v\},\{p\},\{q\},h
\rr
\end{equation}

\begin{eqnarray}
&&
Z_{L\times L}
\ll
u_1-\frac{2i\log(\nome)}{\pi},\ldots,u_L,\{v\},\{p\},\{q\},h
\rr
=
\nonumber \\
&&
\frac{(-)^L}{\textsf{q}^L}\
\exp
\ll
-\pi i \ll Lu_1+(L-2)p_1-\sum_{j=1}^L (v_j+q_j)-2h
\rr
\rr \times \nonumber \\
&& Z_{L\times L}
\ll
\{u\},\{v\},\{p\},\{q\},h
\rr
\end{eqnarray}

\noindent {\it Proof.} Since the weights are entire functions in
the rapidities, it follows that
$Z_{L\times L}\ll\{u\},\{v\},\{p\},\{q\},h\rr$ is an entire function
in $u_1$. To prove the quasi-periodicity conditions, we write the
partition function in the form

\begin{eqnarray}
Z_{L\times L}\ll\{u\},\{v\},\{p\},\{q\},h\rr
&=&
\sum_{n=1}^{L}
P_n\ll u_1,\{v\},p_1,\{q\},h\rr
\times \\
&&
Q_n
\ll 
u_2, \ldots, u_L,\{v\}, p_2, \ldots, p_L, \{q\}, h
\rr
\nonumber
\end{eqnarray}

\noindent where

\begin{eqnarray}
P_n \ll u_1,\{v\},p_1,\{q\},h\rr
&=&
\ll
\prod_{j=1}^{n-1}
a_{+}(u_1,v_j,p_1,q_j)
\rr
c_{+}\ll u_1,v_n,p_1,q_n,h+\sum_{k=1}^{n-1}q_k\rr
\quad\quad\quad
\\
&\times&
\ll
\prod_{j=n+1}^{L}
b_{-}(u_1,v_j,p_1,q_j,h+\sum_{k=1}^{j-1}q_k)
\rr
\nonumber
\end{eqnarray}

\noindent and 
$Q_n \ll u_2, \ldots, u_L, \{v\}, p_2, \ldots, p_L, \{q\},h \rr$
does not depend on $u_1$. Using the expressions for the weights, 
we have

\begin{eqnarray}
&&
P_n\ll u_1+2,\{v\},p_1,\{q\},h\rr
=
(-)^L\ P_n \ll u_1,\{v\},p_1,\{q\},h\rr
\\
&&
P_n\ll u_1-\frac{2i\log(\nome)}{\pi},\{v\},p_1,\{q\},h\rr
=
\\
&&
\frac{(-)^L}{\nome^L}
\exp\ll -\pi i \ll Lu_1+(L-2)p_1-\sum_{j=1}^L (v_j+q_j)-2h\rr\rr
P_n\ll u_1,\{v\},p_1,\{q\},h\rr
\nonumber
\end{eqnarray}

\noindent from which the required property follows immediately.

\subsubsection{Property 2: Simple zeros} The partition function
has simple zeros at $u_1=u_j-p_1-p_j$, where $j=2,\ldots,L$.

\bigskip

\noindent {\it Proof.} We multiply the partition function by
$a_{+}(u_2,u_1,p_2,p_1)$, and use the Yang-Baxter equation to
slide the inserted face through the lattice.

%
\begin{center}
\begin{minipage}{5.0in}
\setlength{\unitlength}{0.0016cm}
\begin{picture}(4800, 5000)(-1000,-0500)
%
\thicklines
\path(0600,3900)(4200,3900)
\path(0900,3300)(4200,3300)
\path(0600,2700)(4200,2700)
\path(0600,2100)(4200,2100)
\path(0600,1500)(4200,1500)
\path(0600,0900)(4200,0900)
\path(0600,0300)(4200,0300)
\path(0600,2700)(0600,0300)
\path(1200,3900)(1200,0300)
\path(1800,3900)(1800,0300)
\path(2400,3900)(2400,0300)
\path(3000,3900)(3000,0300)
\path(3600,3900)(3600,0300)
\path(4200,3900)(4200,0300)
\path(0600,3900)(0900,3300)
\path(0900,3300)(0600,2700)
\path(0600,3900)(0300,3300)
\path(0300,3300)(0600,2700)
\path(0900,3300)(4200,3300)
\thinlines
%
%
\dashline{60.00}(0300,3000)(0450,3000)
\dashline{60.00}(0450,3000)(0750,3600)
\dashline{60.00}(0750,3600)(4500,3600)
\dashline{60.00}(0300,2400)(4500,2400)
\dashline{60.00}(0300,1800)(4500,1800)
\dashline{60.00}(0300,1200)(4500,1200)
\dashline{60.00}(0300,0600)(4500,0600)
\put(4600,3600){$u_1, p_1$}
\put(4600,3000){$u_2, p_2$}
\dashline{60.00}(0300,3600)(0450,3600)
\dashline{60.00}(0450,3600)(0750,3000)
\dashline{60.00}(0750,3000)(4500,3000)
%
\whiten\path(-200,3555)(-020,3600)(-200,3645)(-200,3555)
\path(-380,3600)(-200,3600) 
\put(-1100,3600){$u_2, p_2$}
%
\whiten\path(-200,2955)(-020,3000)(-200,3045)(-200,2955)
\path(-380,3000)(-200,3000) 
\put(-1100,3000){$u_1, p_1$}
%
%
\dashline{60.000}(0900,4200)(0900,0000)
\dashline{60.000}(1500,4200)(1500,0000)
\dashline{60.000}(2100,4200)(2100,0000)
\dashline{60.000}(2700,4200)(2700,0000)
\dashline{60.000}(3300,4200)(3300,0000)
\dashline{60.000}(3900,4200)(3900,0000)
\end{picture}
\begin{ca}
\label{a+}
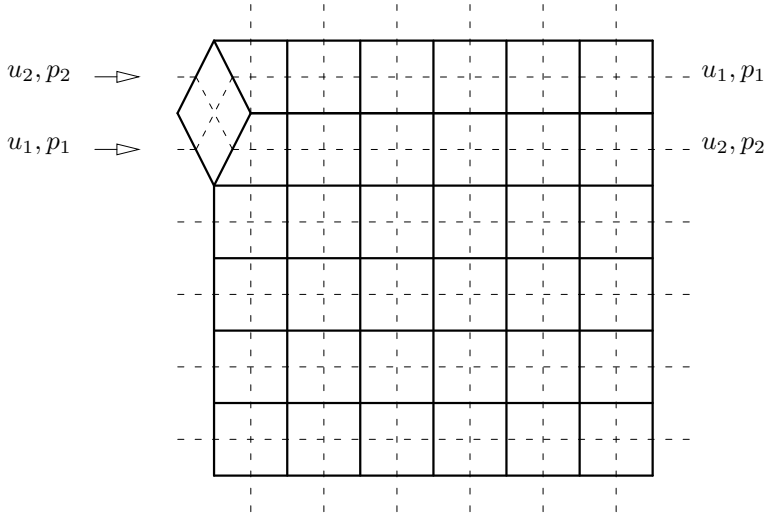
Inserting an $a_{+}$ vertex into the left boundary.
\end{ca}
\end{minipage}
\end{center}
\bigskip

It emerges as $a_{-}(u_2,u_1,p_2,p_1)$, and the order
of the first two lattice rows is reversed.

%
\begin{center}
\begin{minipage}{5.0in}
\setlength{\unitlength}{0.0016cm}
\begin{picture}(4800, 5000)(-1250,-0500)
%
\thicklines
%
%
\thicklines
%
\path(0300,3900)(3900,3900)
\path(0300,3300)(3600,3300)
\path(0300,2700)(3900,2700)
\path(0300,2100)(3900,2100)
\path(0300,1500)(3900,1500)
\path(0300,0900)(3900,0900)
\path(0300,0300)(3900,0300)
%
\path(0300,3900)(0300,0300)
\path(0900,3900)(0900,0300)
\path(1500,3900)(1500,0300)
\path(2100,3900)(2100,0300)
\path(2700,3900)(2700,0300)
\path(3300,3900)(3300,0300)
\path(3900,2700)(3900,0300)
\path(3900,3900)(3600,3300)
\path(3900,3900)(4200,3300)
\path(4200,3300)(3900,2700)
\path(3600,3300)(3902,2700)
\thinlines
%
%
\dashline{60.000}(0100,2400)(4200,2400)
\dashline{60.000}(0100,1800)(4200,1800)
\dashline{60.000}(0100,1200)(4200,1200)
\dashline{60.000}(0100,0600)(4200,0600)
\dashline{60.000}(0100,3600)(3750,3600)
\dashline{60.000}(3750,3600)(4050,3000)
\dashline{60.000}(4050,3000)(4200,3000)
\dashline{60.000}(0100,3000)(3750,3000)
\dashline{60.000}(3750,3000)(4050,3600)
\dashline{60.000}(4050,3600)(4200,3600)
%
%
%
\dashline{60.000}(0600,4200)(0600,0000)
\dashline{60.000}(1200,4200)(1200,0000)
\dashline{60.000}(1800,4200)(1800,0000)
\dashline{60.000}(2400,4200)(2400,0000)
\dashline{60.000}(3000,4200)(3000,0000)
\dashline{60.000}(3600,4200)(3600,0000)
%
\whiten\path(-300,3555)(-120,3600)(-300,3645)(-300,3555)
\path(-480,3600)(-300,3600) 
\put(-1100,3600){$u_2, p_2$}
\put(+4300,3660){$u_1, p_1$}
%
\whiten\path(-300,2955)(-120,3000)(-300,3045)(-300,2955)
\path(-480,3000)(-300,3000) 
\put(-1100,3000){$u_1, p_1$}
\put(+4300,3000){$u_2, p_2$}
\end{picture}
\begin{ca}
\label{a-}
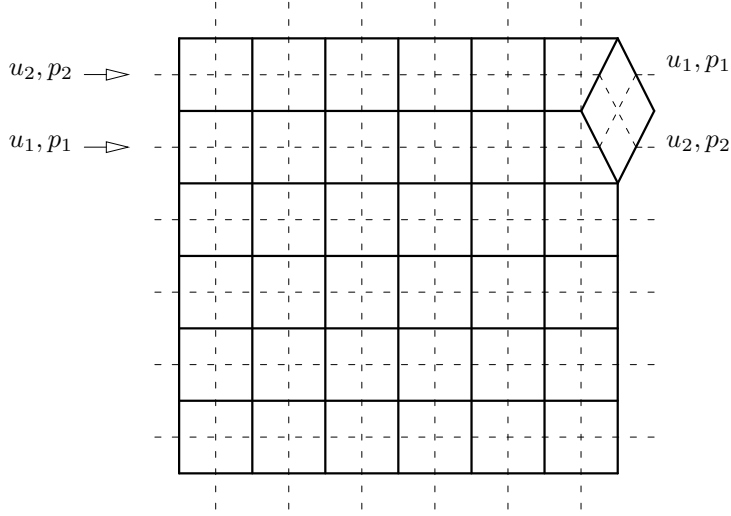
Extracting an $a_{-}$ vertex from the right boundary.
\end{ca}
\end{minipage}
\end{center}
\bigskip

This is equivalent to the equation
\begin{eqnarray}
Z_{L\times L}\ll\{u\},\{v\},\{p\},\{q\},h\rr
&=&
\frac{
a_{-}(u_2,u_1,p_2,p_1)
}
{
a_{+}(u_2,u_1,p_2,p_1)
}
\times \\
&&
Z_{L\times L}
\ll
u_2, u_1, \ldots, u_L, \{v\}, p_2, p_1, \ldots, p_L, \{q\}, h
\rr
\nonumber
\end{eqnarray}

Repeating this procedure on the second and third rows, and so
on, we obtain

\begin{eqnarray}
Z_{L\times L}\ll\{u\},\{v\},\{p\},\{q\},h\rr
&=&
\prod_{j=2}^{L}
\ll
\frac{
a_{-}(u_j,u_1,p_j,p_1)
}
{
a_{+}(u_j,u_1,p_j,p_1)
}
\rr
\times \\
&&
Z_{L\times L}
\ll 
u_2, \ldots, u_L, u_1, \{v\}, p_2, \ldots, p_L, p_1, \{q\}, h
\rr
\nonumber
\end{eqnarray}

\noindent which has the required simple zeros in the numerator.

\subsubsection{Property 3: A recursion relation} The partition
function satisfies the recursion relation

\begin{eqnarray}
&&
\left.
Z_{L\times L}
\ll
\{u\},\{v\},\{p\},\{q\},h
\rr\right|_{u_1= v_1-p_1-q_1}
=
c_{+}(v_1-p_1-q_1,v_1,p_1,q_1,h) \times
\quad\quad
\nonumber \\
&&
\ll
\prod_{j=2}^{L}
b_{+}(u_j,v_1,p_j,q_1,h+\sum_{k=1}^{j-1}p_k)\
b_{-}(v_1-p_1-q_1,v_j,p_1,q_j,h+\sum_{k=1}^{j-1}q_k)
\rr
\times
\nonumber
\\
&&
Z_{(L-1)\times (L-1)}
\ll
u_2,\ldots,u_L,v_2,\ldots,v_L,p_2,\ldots,p_L,q_2,\ldots,q_L,
h + p_1 + q_1 - 1
\rr
\end{eqnarray}

\noindent {\it Proof.} In any lattice configuration in the partition
function sum, the top-left corner of the lattice must be
$a_{+}(u_1,v_1,p_1,q_1)$ or $c_{+}(u_1,v_1,p_1,q_1, h)$. Setting
$u_1=v_1-p_1-q_1$ in the partition function sets to zero all
configurations with $a_{+}(u_1,v_1,p_1,q_1)$.

The surviving configurations must have a top-left corner equal to $c_{+}$,
which fixes the rest of the top row to $b_{-}$, the rest of the first
column to $b_{+}$, and the remainder of the lattice to $Z_{(L-1)\times
(L-1)}$. The above recursion follows immediately from these
considerations.

\subsubsection{Property 4} The partition function on a $1\times 1$
lattice is given by

\begin{eqnarray}
Z_{1\times 1}(u_1,v_1,p_1,q_1,h)
=
c_{+}(u_1,v_1,p_1,q_1,h)
\end{eqnarray}
\noindent

\noindent {\it Proof.} This follows from the definition of domain wall
boundary conditions.

\subsection{The partition function is uniquely determined}

Assume that $Z_{(n-1) \times (n-1)}$ is uniquely determined by
the above four properties, for some $n\geq 2$. From Property
{\bf 1}, Property {\bf 2} and Theorem {\bf 1.1}, we have

\begin{eqnarray}
&&
Z_{n\times n}
\ll
\{u\},\{v\},\{p\},\{q\}, h
\rr = \kappa(u_2, \ldots, u_n,\{v\},\{p\},\{q\}) \times
\nonumber
\\
&&
[\sum_{j=1}^{n}(v_j-u_j)+\sum_{j=1}^{n}(p_j+q_j)+2h]
 \ll \prod_{j=2}^{n}[u_1-u_j+p_1+p_j]\rr
\end{eqnarray}

Property {\bf 3} fully determines the coefficient $\kappa$ in terms
of $Z_{(n-1)\times (n-1)}$. Finally, since $Z_{1\times 1}$ is uniquely
determined by Property {\bf 4}, $Z_{n\times n}$ is uniquely determined
by the four properties.

\subsection{The domain wall partition function}

The solution to the preceding four properties is given by

\bigskip

\begin{boxedminipage}[l]{14cm}
\begin{equation}
\begin{split}
&
Z_{L\times L}
\ll
\{u\},\{v\},\{p\},\{q\},h
\rr
=
\frac{
\prod_{j=1}^{L}
[2p_j]^{\frac{1}{2}}
[2q_j]^{\frac{1}{2}}
}
{
{[2(h+\sum_{j=1}^{L}p_j)]}^{\frac{1}{2}}
{[2(h+\sum_{j=1}^{L}q_j)]}^{\frac{1}{2}}
} \times 
{}\\
{}&{}
[
\sum_{j=1}^{L}(v_j-u_j) +
\sum_{j=1}^{L}(p_j+q_j) + 2h
]
\prod_{1\leq j < k \leq L}
[u_j-u_k+p_j+p_k]
[v_k-v_j+q_k+q_j]
\end{split}
\end{equation}
\end{boxedminipage}
\bigskip

\subsection{Two choices of DWBC's} There are two possible choices 
of DWBC's. One corresponds to an {\it expanded} $c_{+}$, as in this 
work, and one to an `expanded' $c_{-}$ vertex. The DWPF depends on 
the choice. The two expressions coincide for vanishing external
fields, that is $p_i = q_j = \frac{1}{2}$, and appropriate choices 
of the boundary height variables.

\section{A Perk-Schultz-type $gl(1|1)$ height model}

In \cite{dm}, Deguchi and Martin introduced elliptic height versions 
of the $gl(r+1|s+1)$ trigonometric vertex models. In the following, 
we define DWBC's and compute the DWPF in the $gl(1|1)$ case. 
Since the analysis in this Section follows almost verbatim that of 
Section {\bf 2}, we will be brief and give just enough details where 
the two models differ. 

\subsection{Notation, heights and restrictions}

In this model, there are two square lattices. A {\it physical} 
$L\times L$ lattice that the heights live on, and a {\it target} 
$\ZZ\times \ZZ$ lattice that the heights take values in.
The target lattice is spanned by the unit vectors $\hat{e}_{\mu}$, 
$\mu \in \{-1, +1\}$, thus the height variables are 2-component 
vectors $\{h_{\mu}, h_{\nu}\}$. Heights on adjacent physical lattice 
corners are restricted to take values in adjacent points on the 
target lattice. 
Height differences along $\hat{e}_{-1}$ and $\hat{e}_{+1}$ lead 
to different vertex weights. To each height vector 
$h= \{h_{\mu}, h_{\nu}\}$, we assign a scalar

\begin{equation}
h_{\mu \nu} = h_{\mu} + h_{\nu} + \omega_{\mu \nu}
\end{equation}

\noindent where $\omega$ is an arbitrary constant antisymmetric
complex $2\times 2$ matrix. We use $\hat{e}_{\mu}$ as well as 
$\hat{e}_{\textrm{sign($\mu$)}}$ to indicate the same unit vector.

\subsection{No external fields and the crossing parameter is 
a variable} Unlike the previous Felderhof-type model, the 
Perk-Schultz-type model in this Section has no external fields. 
The crossing parameter is left as a variable.

\subsection{The weights} 

The non-zero vertex weights are

\begin{equation}
W_{uv}
\ll
\begin{array}{ll}
h                & h+   \hat{e}_{\mu} \\
h+ \hat{e}_{\mu} & h+ 2 \hat{e}_{\mu}
\end{array}
\rr
= a_{\mu}(u, v)
= \frac{[ 1+ \mu (u-v) ]}{[1]}
\label{a-ps}
\end{equation}

\begin{equation}
W_{uv}
\ll
\begin{array}{ll}
h                & h+ \hat{e}_{\nu} \\
h+ \hat{e}_{\mu} & h+ \hat{e}_{\mu}+ \hat{e}_{\nu}
\end{array}
\rr
= b_{\mu}(u, v)
= \frac{[u-v][h_{\mu\nu}-1]}{[1][h_{\mu\nu}]},
\quad
\mu \ne \nu
\label{b-ps}
\end{equation}

\begin{equation}
W_{uv}
\ll
\begin{array}{ll}
h                & h+ \hat{e}_{\mu} \\
h+ \hat{e}_{\mu} & h+ \hat{e}_{\mu}+ \hat{e}_{\nu}
\end{array}
\rr
= c_{\mu}(u, v)
= \frac{[h_{\mu\nu}- (u-v)]}{[h_{\mu\nu}]},
\quad
\mu \ne \nu
\label{c-ps}
\end{equation}

The variables $h_{\mu \nu}$ on the right hand sides of Equations 
{\bf \ref{b-ps}} and {\bf \ref{c-ps}} are the scalars assigned to 
the heights at the upper left corners of the corresponding vertices.

\subsection{The Yang-Baxter equations}

The weights satisfy the following Yang-Baxter equations
\cite{dm}.

\begin{eqnarray}
&&\sum_{g \in \field{Z}^2}
W_{u_1 u_2}
\ll
\begin{array}{ll}
b & g \\
a & f
\end{array}
\rr
W_{u_1 u_3}
\ll
\begin{array}{ll}
g & d \\
f & e
\end{array}
\rr
W_{u_2 u_3}
\ll
\begin{array}{ll}
b & c \\
g & d
\end{array}
\rr
= \nonumber \\
&&\sum_{g \in \field{Z}^2}
W_{u_2 u_3}
\ll
\begin{array}{ll}
a & g \\
f & e
\end{array}
\rr
W_{u_1 u_3}
\ll
\begin{array}{ll}
b & c \\
a & g
\end{array}
\rr
W_{u_1 u_2}
\ll
\begin{array}{ll}
c & d \\
g & e
\end{array}
\rr
\end{eqnarray}

\subsection{The $c_{+}$ vertex}

We take the $c_{+}$ vertex to be as in Figure {\bf \ref{c++}}.

%
\begin{center}
\begin{minipage}{3.0in}
\setlength{\unitlength}{0.0016cm}
\begin{picture}(3000,2600)(0,300)
%
\thicklines
%
\path(1500,2400)(2700,2400)
\path(1500,1200)(2700,1200)
%
\path(1500,2400)(1500,1200)
\path(2700,2400)(2700,1200)
%
\dashline{60.00}(1200,1800)(3000,1800)
%
\whiten\path(0840,1740)(1000,1800)(0840,1860)(0840,1740)
\path(0680,1800)(0840,1800)
 \put(0500,1900){\u}
%
\dashline{60.00}(2100,2700)(2100,0900)
%
\whiten\path(2040,0540)(2100,0700)(2160,0540)(2040,0540)
%
\path(2100,0540)(2100,0380)
\put(2200,0380){\v}
\put(1500,2400){\blacken\ellipse{050}{050}}
\put(1300,2500){\h}
\put(1500,1200){\blacken\ellipse{050}{050}}
\put(1300,1000){\h\+\textit{\scriptsize{$\hat{e}_{+}$}}}
\put(2700,2400){\blacken\ellipse{050}{050}}
\put(2800,2500){\h\+\textit{\scriptsize{$\hat{e}_{+}$}}}
\put(2700,1200){\blacken\ellipse{050}{050}}
\put(2800,1000){\h\+\textit{\scriptsize{$\hat{e}_{+}$\+$\hat{e}_{-}$}}}
\end{picture}
\begin{ca}
\label{c++}
A general face configuration.
\end{ca}
\end{minipage}
\end{center}
\bigskip

%
\begin{center}
\begin{minipage}{5.0in}
\setlength{\unitlength}{0.0016cm}
\begin{picture}(4800, 5200)(-2000, -1200)
\thicklines
%
%
\path(0000,3600)(3600,3600)
\path(0000,3000)(3600,3000)
\path(0000,2400)(3600,2400)
\path(0000,1800)(3600,1800)
\path(0000,1200)(3600,1200)
\path(0000,0600)(3600,0600)
\path(0000,0000)(3600,0000)
%
%
\path(0000,3600)(0000,0000)
\path(0600,3600)(0600,0000)
\path(1200,3600)(1200,0000)
\path(1800,3600)(1800,0000)
\path(2400,3600)(2400,0000)
\path(3000,3600)(3000,0000)
\path(3600,3600)(3600,0000)
\put(0000,3600){\blacken\circle{50,50}}
\put(3600,3600){\blacken\circle{50,50}}
\put(0000,3000){\blacken\circle{50,50}}
\put(0000,2400){\blacken\circle{50,50}}
\put(0000,1800){\blacken\circle{50,50}}
\put(0000,1200){\blacken\circle{50,50}}
\put(0000,0600){\blacken\circle{50,50}}
\put(-400,3600){$h                $}
\put(-900,3000){$h +   \hat{e}_{+}$}
\put(-900,2400){$h + 2 \hat{e}_{+}$}
\put(-900,0000){$h + L \hat{e}_{+}$}
\put(3700,3600){$h + L \hat{e}_{+}$}
\put(3700,3000){$h + L \hat{e}_{+} +   \hat{e}_{-}$}
\put(3700,2400){$h + L \hat{e}_{+} + 2 \hat{e}_{-}$}
\put(3700,0000){$h + L \hat{e}_{-} + L \hat{e}_{+}$}

\thicklines
\blacken\path(0555,-500)(0645,-500)(0600,-320)(0555,-500)
\path(0600,-980)(0600,-500)
\put(0600,-1200){$h + L \hat{e}_{+} +   \hat{e}_{-}$}

\thicklines
\blacken\path(1155,-500)(1245,-500)(1200,-320)(1155,-500)
\path(1200,-680)(1200,-500)
\put(1200,-0900){$h + L \hat{e}_{+} + 2 \hat{e}_{-}$}

\put(0000,0000){\blacken\circle{50,50}}
\put(0600,0000){\blacken\circle{50,50}}
\put(3600,0000){\blacken\circle{50,50}}

\thinlines
%
%
\dashline{60.00}(-200,3300)(3900,3300)
\dashline{60.00}(-200,2700)(3900,2700)
\dashline{60.00}(-200,2100)(3900,2100)
\dashline{60.00}(-200,1500)(3900,1500)
\dashline{60.00}(-200,0900)(3900,0900)
\dashline{60.00}(-200,0300)(3900,0300)
%
%
\dashline{60.00}(0300,3900)(0300,-200)
\dashline{60.00}(0900,3900)(0900,-200)
\dashline{60.00}(1500,3900)(1500,-200)
\dashline{60.00}(2100,3900)(2100,-200)
\dashline{60.00}(2700,3900)(2700,-200)
\dashline{60.00}(3300,3900)(3300,-200)
\end{picture}
\begin{ca}
\label{dwbcps}
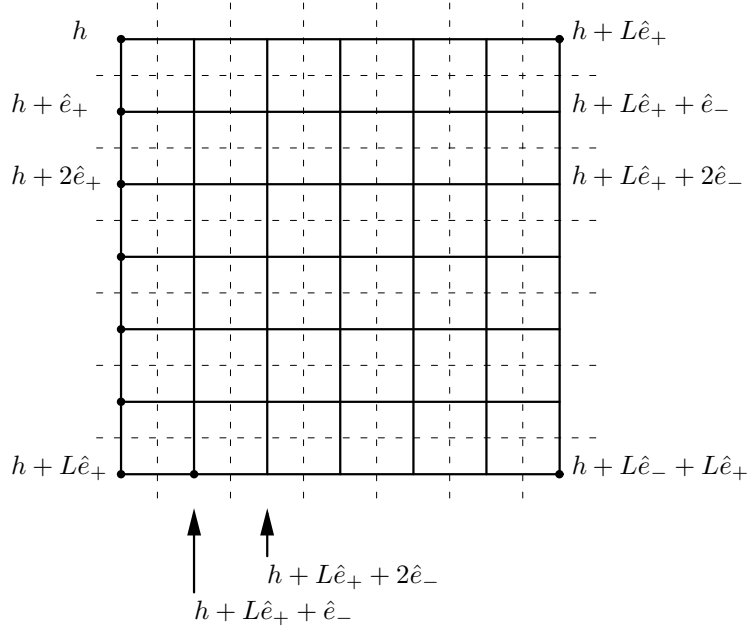
Perk-Schultz-type height domain wall boundary conditions.
\end{ca}
\end{minipage}
\end{center}
\bigskip

\subsection{The domain wall boundary conditions}
We choose the DWBC as in Figure {\bf \ref{dwbcps}}.
In other words, starting from the lower left corner,
all height changes along the left boundary are of type
$\hat{e}_{+1}$, along the upper boundary they are of type
$\hat{e}_{-1}$, along the right boundary they are of type
$\hat{e}_{+1}$, then along the lower boundary they are of
type $\hat{e}_{-1}$.

\subsection{The line-permuting vertices}
We take the line permuting vertices to be as in Figure
{\bf \ref{lpps}}. Their weights have different zeros leading 
to a factorization of the DWPF, just as in Section {\bf 2}.
%
%
\begin{center}
\begin{minipage}{5.0in}
\setlength{\unitlength}{0.0016cm}
\begin{picture}(3000,2800)(0,100)
%
\thicklines
%
\path(1500,2400)(2700,2400)
\path(1500,1200)(2700,1200)
%
\path(1500,2400)(1500,1200)
\path(2700,2400)(2700,1200)
\thinlines
\dashline{60.00}(1200,1800)(3000,1800)
\thicklines
%
\whiten\path(0840,1740)(1000,1800)(0840,1860)(0840,1740)
\path(0680,1800)(0840,1800)
 \put(0500,1900){\u}
\thinlines
\dashline{60.00}(2100,2700)(2100,0900)
\thicklines
%
\whiten\path(2040,0540)(2100,0700)(2160,0540)(2040,0540)
%
\path(2100,0540)(2100,0380)
\put(2200,0380){\v}
\put(1500,2400){\blacken\ellipse{050}{050}}
\put(1300,2500){\h}
\put(1500,1200){\blacken\ellipse{050}{050}}
\put(1300,1000){\h\+\textit{\scriptsize{$\hat{e}_{+}$}}}
\put(2700,2400){\blacken\ellipse{050}{050}}
\put(2500,2500){\h\+\textit{\scriptsize{$\hat{e}_{+}$}}}
\put(2700,1200){\blacken\ellipse{050}{050}}
\put(2500,1000){\h\+\2\textit{\scriptsize{$\hat{e}_{+}$}}}
%
%
\path(4500,2400)(5700,2400)
\path(4500,1200)(5700,1200)
%
\path(4500,2400)(4500,1200)
\path(5700,2400)(5700,1200)
%
\thinlines
\dashline{60.00}(4200,1800)(6000,1800)
\thicklines
%
\whiten\path(3840,1740)(4000,1800)(3840,1860)(3840,1740)
\path(3680,1800)(3840,1800)
 \put(3500,1900){\u}
%
\thinlines
\dashline{60.00}(5100,2700)(5100,0900)
\thicklines
%
\whiten\path(5040,0540)(5100,0700)(5160,0540)(5040,0540)
%
\path(5100,0540)(5100,0380)
 \put(5200,0380){\v}
\put(4500,2400){\blacken\ellipse{050}{050}}
\put(4300,2500){\h}
\put(4500,1200){\blacken\ellipse{050}{050}}
\put(4300,1000){\h\+\textit{\scriptsize{$\hat{e}_{-}$}}}
\put(5700,2400){\blacken\ellipse{050}{050}}
\put(5500,2500){\h\+\textit{\scriptsize{$\hat{e}_{-}$}}}
\put(5700,1200){\blacken\ellipse{050}{050}}
\put(5500,1000){\h\+\2\textit{\scriptsize{$\hat{e}_{-}$}}}
\end{picture}
\begin{ca}
\label{lpps}
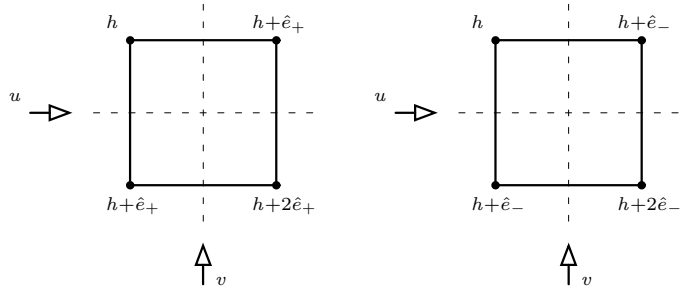
The Perk-Schultz-type line permuting vertices
$a_{+}$ and $a_{-}$.
\end{ca}
\end{minipage}
\end{center}
\bigskip
\subsection{The DWPF}

Having defined the model, the derivation of the corresponding
DWPF proceeds precisely in analogy with that in Section {\bf 2}.
Based on the quasi-periodicity properties of the partition
function, we propose a factorization in terms of the $[u]$
functions. We obtain the zeros by permuting adjacent flow
lines, using the line permuting vertices, obtain a recursion
relation the DWPF satisfies and an initial condition. The DWPF 
is uniquely determined, and the following expression satisfies
all the conditions.

\bigskip
\begin{boxedminipage}[l]{12cm}
\begin{equation}
\begin{split}
&Z^{}_{L \times L} = \\
&\frac{[h +(L-1) - \sum^L_{k=1}(u_k-v_k)]}
      {[h +(L-1)]}
\prod_{1\le i < j \le L}
\frac{[1+u_i-u_j]}{[1]} \frac{[1-(v_i-v_j)]}{[1]}
\label{mas}
\end{split}
\end{equation}
\end{boxedminipage}
\bigskip

\subsection{On the $gl(r+1|s+1)$ height models} The reason
we chose the $gl(1|1)$ case is that, given the way that we define
DWBC's, the symmetries of the $gl(r+1|s+1)$ models are such that
only the two state variables variables that we put on the domain
wall boundaries end up propagating inside the configurations.
This effectively restricts the DWPF to the $gl(1|1)$ model.

\section{Remarks}

The point of this work is to give examples of domain wall partition 
functions in elliptic and/or height models. We restricted our attention 
to models that are non-invariant under state variable conjugation, which 
greatly simplified the problem.

Because of the fermionic nature of the models discussed in this paper, 
there is no interesting applications of their DWPF's to enumerations 
of alternating sign matrices or related objects that we are aware of.

It is highly likely that all models that are non-invariant under some 
form of state variable conjugation, that allows factorization, are 
fermionic (as the Felderhof-type model of Section {\bf 2}) or contain 
fermions that play an essential role in the definition of the DWBC's
(as the Perk-Schultz-type model of Section {\bf 3}). As such, these 
models are unrepresentative of the general case. However, they are 
non-trivial and we hope that one can learn something by extending our 
results to compute correlation functions.

\section*{Acknowledgements}
OF would like to thank Professors R~J~Baxter, T~Deguchi and M~Jimbo
for discussions, and P~Bouwknegt, T~Guttmann and T~Miwa for hospitality 
at ANU, MASCOS and Kyoto University, while this work was in progress. 
MW and MZ are supported by an Australian Postgraduate Award (APA).

\end{document}